\def   \ni {\noindent}

\def   \ssk {\vskip  5truept}

\def   \bsk {\vskip 15truept}

\def   \newline {\hfil\break}

\newcommand{\simgt}{\mbox{$\stackrel{>}{_{\sim}}$} }

\newcommand{\Omm}{\Omega_{\rm m}}
\newcommand{\Oml}{\Omega_{\Lambda}}
\def\etal{{\frenchspacing\it et al.}}
\newcommand{\mnras}{{MNRAS}}
\newcommand{\apj}{{ApJ}}
\newcommand{\apjl}{{ApJL}}
\newcommand{\apjs}{{ApJS}}
\newcommand{\nat}{{Nature}}
\def\eg{{\frenchspacing\it e.g. }}

\documentstyle[epsfig]{article}
\begin{document}

%
\def\la{\mathrel{\mathchoice {\vcenter{\offinterlineskip\halign{\hfil
$\displaystyle##$\hfil\cr<\cr\sim\cr}}}
{\vcenter{\offinterlineskip\halign{\hfil$\textstyle##$\hfil\cr
<\cr\sim\cr}}}
{\vcenter{\offinterlineskip\halign{\hfil$\scriptstyle##$\hfil\cr
<\cr\sim\cr}}}
{\vcenter{\offinterlineskip\halign{\hfil$\scriptscriptstyle##$\hfil\cr
<\cr\sim\cr}}}}}
\def\ga{\mathrel{\mathchoice {\vcenter{\offinterlineskip\halign{\hfil
$\displaystyle##$\hfil\cr>\cr\sim\cr}}}
{\vcenter{\offinterlineskip\halign{\hfil$\textstyle##$\hfil\cr
>\cr\sim\cr}}}
{\vcenter{\offinterlineskip\halign{\hfil$\scriptstyle##$\hfil\cr
>\cr\sim\cr}}}
{\vcenter{\offinterlineskip\halign{\hfil$\scriptscriptstyle##$\hfil\cr
>\cr\sim\cr}}}}}
\def\degr{\hbox{$^\circ$}}
\def\arcmin{\hbox{$^\prime$}}
\def\arcsec{\hbox{$^{\prime\prime}$}}

\hsize 5truein
\vsize 8truein
\font\abstract=cmr8
\font\keywords=cmr8
\font\caption=cmr8
\font\references=cmr8
\font\text=cmr10
\font\affiliation=cmssi10
\font\author=cmss10
\font\mc=cmss8
\font\title=cmssbx10 scaled\magstep2
\font\alcit=cmti7 scaled\magstephalf
\font\alcin=cmr6 
\font\ita=cmti8
\font\mma=cmr8
\def\ref{\par\noindent\hangindent 15pt}
\null


\title{\ni Combining Supernovae and LSS Information with the CMB
}                                               

\bsk \bsk
\author{\ni A.N. Lasenby, S.L. Bridle \& M.P. Hobson}
\bsk
\affiliation{ Astrophysics Group, Cavendish Laboratory, Madingley
Road, Cambridge, CB3 0HE, UK.
}                                                
\bsk
\baselineskip = 12pt

\abstract{ABSTRACT \ni
Observations of the Cosmic Microwave Background (CMB), large scale
structure (LSS) and standard candles such as Type 1a Supernovae (SN) each
place different constraints on the values of cosmological parameters.
We assume an inflationary Cold Dark Matter model with a cosmological
constant, in which the initial density perturbations in the universe
are adiabatic. 
We discuss the parameter degeneracies inherent in interpreting CMB or
SN data, and derive their orthogonal nature. We then present our
preliminary results of combining CMB and SN likelihood functions. 
The results of combining the CMB and IRAS 1.2 Jy survey information
are given, with marginalised confidence regions in the $H_0$,
$\Omega_m$, $b_{\rm{IRAS}}$ and $Q_{\rm{rms-ps}}$ directions 
assuming $n=1$, $\Omega_{\Lambda}+\Omega_m=1$ and $\Omega_b
h^2=0.024$. 
Finally we combine all three likelihood functions and find that the
three data sets are consistent and suitably orthogonal, leading to
tight constraints on $H_0$, $\Omega_m$, $b_{\rm{IRAS}}$ and
$Q_{\rm{rms-ps}}$, given our assumptions. 
}                                                    
\bsk
\baselineskip = 12pt

\bsk
\baselineskip = 12pt

\text{\ni 1. INTRODUCTION
\ssk
\ni     

By comparing the observed CMB power spectrum with predictions from
cosmological models one can estimate cosmological parameters. This has
become an area of great current interest, with many groups carrying
out the analyses for a range of assumed models 
(\eg Hancock \etal~1998, Lineweaver \etal~1997, Bond \&
Jaffe1998). Generally speaking, the results of using CMB data alone to
do this are broadly consistent with the expected range of cosmological
parameters, though perhaps with a tendency for $H_0$ to come out
rather low (assuming spatially flat models). In an independent manner,
similar predictions can be made by comparing Large Scale Structure
(LSS) surveys with cosmological models~(Willick
\etal~1997, Fisher \& Nusser 1996, Heavens \& Taylor 1995). Also, when
distant Type 1A supernovae are used as a standard candle, one can
assess the probability as a function of
$\Omega_m$ and $\Omega_{\Lambda}$. We demonstrate that the results
from the different data sets are compatible. However using each data
set alone it is only possible to constrain combinations of the
cosmological parameters. We show how these combinations of parameters
are different for each data set, therefore when we combine the data
sets we obtain tight constraints.   

\bsk
\ni 2. THE COMPLEMENTARY NATURE OF SUPERNOVAE AND CMB DATA
\ssk
\ni 

%
%
\normalsize{
\begin{figure}[t]
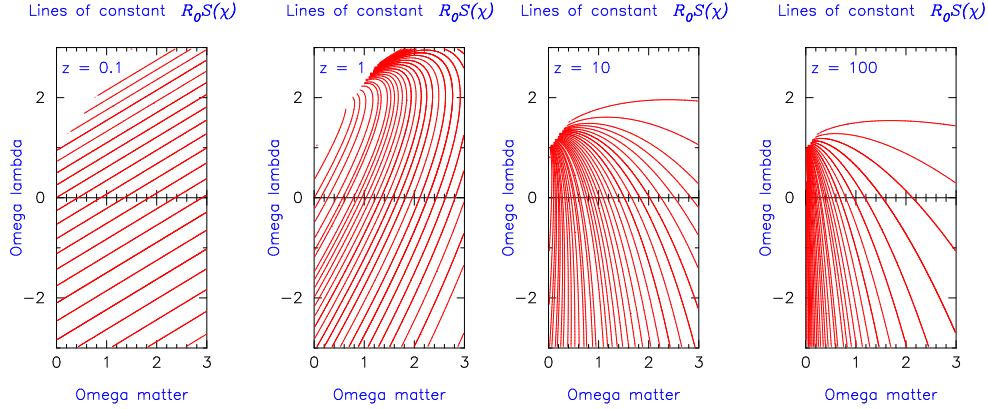

\centerline{
\begin{tabular}{cccc}
\mbox{\psfig{figure=santander_plot_z_0p1.ps,angle=-90,width=3cm}} 
& 
\mbox{\psfig{figure=santander_plot_z_1.ps,angle=-90,width=3cm}} 
\mbox{\psfig{figure=santander_plot_z_10.ps,angle=-90,width=3cm}} 
& 
\mbox{\psfig{figure=santander_plot_z_100.ps,angle=-90,width=3cm}} 
\end{tabular}
}
\caption{\center{FIGURE 1  Plots showing contours of constant $R_0 S(\chi)$ at
redshifts of 0.1, 1, 10 and 100.}}
\end{figure}
}
There has recently been great interest in combining Type Ia
supernovae (SN) data with results from the CMB (e.g. 
Lineweaver 1998, Tegmark 1998). It is 
instructive to see how the complementarity between the supernovae and
CMB data arises.
The key quantity for this discussion is $R_0 S(\chi)$,
which occurs in the definitions of {\em Luminosity Distance}:
\begin{displaymath}
d_L = R_0 S(\chi) (1+z),
\end{displaymath}
and {\em Angular Diameter Distance:} 
\begin{displaymath}
d_{\theta} = R_0 S(\chi)/(1+z).
\end{displaymath}
Here $R_0$ is the current scale factor of the Universe, $\chi$ is a
comoving coordinate, and $S(\chi)$ is $\sinh(\chi)$, $\chi$ or
$\sin(\chi)$ depending on whether the universe is open, flat or closed
respectively. For a general Friedmann-Lemaitre model, one finds that
\begin{displaymath}
R_0 S(\chi) \propto \frac{1}{|\Omega_{\rm k}|^{1/2}} {\rm sin(h)} \left\{ 
|\Omega_{\rm k}|^{1/2} \int_0^z \frac{dz'}{H(z')} \right\}
\end{displaymath}
where
\begin{eqnarray*}
\Omega_{\rm k} & = & 1 - (\Omm + \Oml), \\
H^2(z) & = & H_0^2 \left( (1+\Omm z)(1+z)^2 - \Oml z (2+z) \right).
\end{eqnarray*}
For small $z$, it is easy to show that
\begin{displaymath}
d_L \propto z + \frac{1}{2}(1-2q_0)z^2,
\end{displaymath}
where $q_0 = \frac{1}{2}(\Omm - 2\Oml)$ is the usual
deceleration parameter.

Therefore, for small $z$, SN results are degenerate along a line
of constant $q_0$.  However, the contours of equal $R_0 S(\chi)$
shift around as $z$ increases and for $z \simgt 100$ the contours
are approximately orthogonal to those corresponding to $q_0$
constant.  This is the reason why CMB and SN results are
ideally complementary. The current microwave background data is
mainly significant in delimiting the left/right position of the
first Doppler peak in the power spectrum, and this depends on the
cosmology via the angular diameter distance formula, evaluated at
$z \sim 1000$.  Thus the CMB results will tend to be degenerate
along lines roughly perpendicular to those for the supernovae in
the $(\Omm,\Oml)$ plane.

\begin{figure}[t]
\centerline{\psfig{file=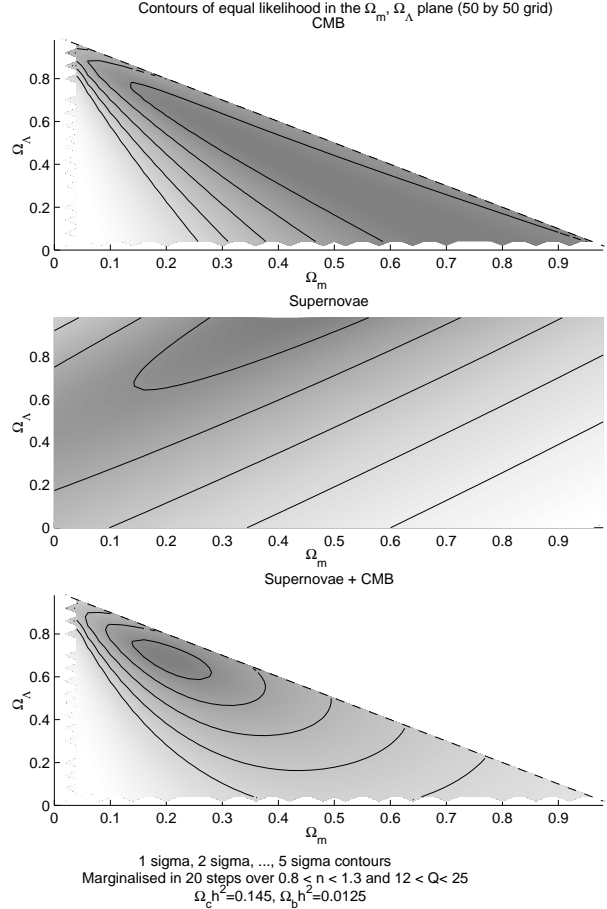,width=8cm}}
\caption{
} 
\label{fig:cmb-sn}
\caption{FIGURE 2  Likelihoods in the $\Omega_m$, $\Omega_{\Lambda}$
plane from CMB and SN data}
\end{figure}
We may calculate the likelihood of a given set of cosmological
parameters for CMB data alone using the bandpower approach described
in \eg  Hancock \etal~1998. We use Seljak and
Zaldariagga's CMBFAST code to calculate scalar mode CMB
power spectra for an adiabatic inflationary Cold Dark Matter universe
and use the CMB data points described in Webster
\etal~1998. In this preliminary analysis we fix 
$\Omega_c h^2 = 0.145$ and $\Omega_b h^2 = 0.0125$ (where $h=H_0/100$
and $\Omega_c=\Omega_m-\Omega_b$). We marginalise 
over the initial power spectrum index, $n$, and CMB power spectrum
normalisation, $C_2$. The result is shown in the top panel of
Figure~2. Contours are at 
changes in $-\log{(\rm{Likelihood})}$ of 0.5, 2, 4.5, 8 and 12.5 from
the minimum value. As expected, there is a degeneracy in a similar
direction to the last plot in Figure 1. 
The SN likelihoods we use are based on data in Perlmutter \etal
~1998 and are plotted in the second panel of
Figure~2. Again, the direction of degeneracy is along the lines of the
first two plots of Figure~1.

The lines of degeneracy are orthogonal and overlap, leaving a small
patch of parameter space that fits both data sets. However we may be more
quantitative than this. The probability of the Universe
having a particular $\Omega_m$, $\Omega_{\Lambda}$ given \emph{both}
the CMB and SN data sets is simply found by multiplying together the
probabilities of those $\Omega_m$, $\Omega_{\Lambda}$ for each data
set. We thereby obtain the final plot in Figure 2. The preferred
universe is close to flat with a low $\Omega_m$. However, we have
applied very limiting assumptions. Detailed likelihood
calculations covering a much wider range of assumptions using both
supernovae and CMB data are currently being 
carried out by Efstathiou \etal, and should be submitted shortly.

\bsk
\ni 3. COMBINING CMB AND LSS
\ssk
\ni 

Recently, Webster \etal~have combined the CMB
likelihoods with IRAS likelihoods obtained from the 1.2Jy galaxy
redshift survey following the spherical harmonic approach of Fisher,
Scharf \& Lahav. They assume a Harrison--Zel'dovich
primordial scalar power spectrum ($n_s=1$) and the nucleosynthesis constraint
$\Omega_b h^2=0.024$ (Tytler, Fan \& Burles 1996)
in a flat universe with a cosmological constant, although clearly it
would be interesting to relax these constraints.  This approach is
complementary to that of Gawiser \& Silk, who used
a compilation of large scale structure and CMB data to assess the
goodness of fit of a wide variety of cosmological models.

Because the CMB and LSS predictions are degenerate with respect to
different parameters (roughly: $\Omega_m {\rm
vs}~\Omega_{\Lambda}$ for CMB; $H_0$ and $\Omega_m {\rm vs}~b_{\rm
iras}$ for LSS), the combined data likelihood analysis allows the
authors to break these degeneracies, giving new parameter
constraints. The different degeneracy directions in the $\Omega_m$,
$h$ plane are shown in the top two panels of Figure~4. Note that the
two data sets agree well in the region where the lines of degeneracy
cross. Figure~3 shows the final 1-dimensional
probability distributions for the main cosmological parameters
after marginalizing over each of the others. The vertical dashed lines
denote the 68\% confidence limits and the horizontal plot limits are at
the 99\% confidence limits. 

\begin{figure}[t]
\centerline{\psfig{file=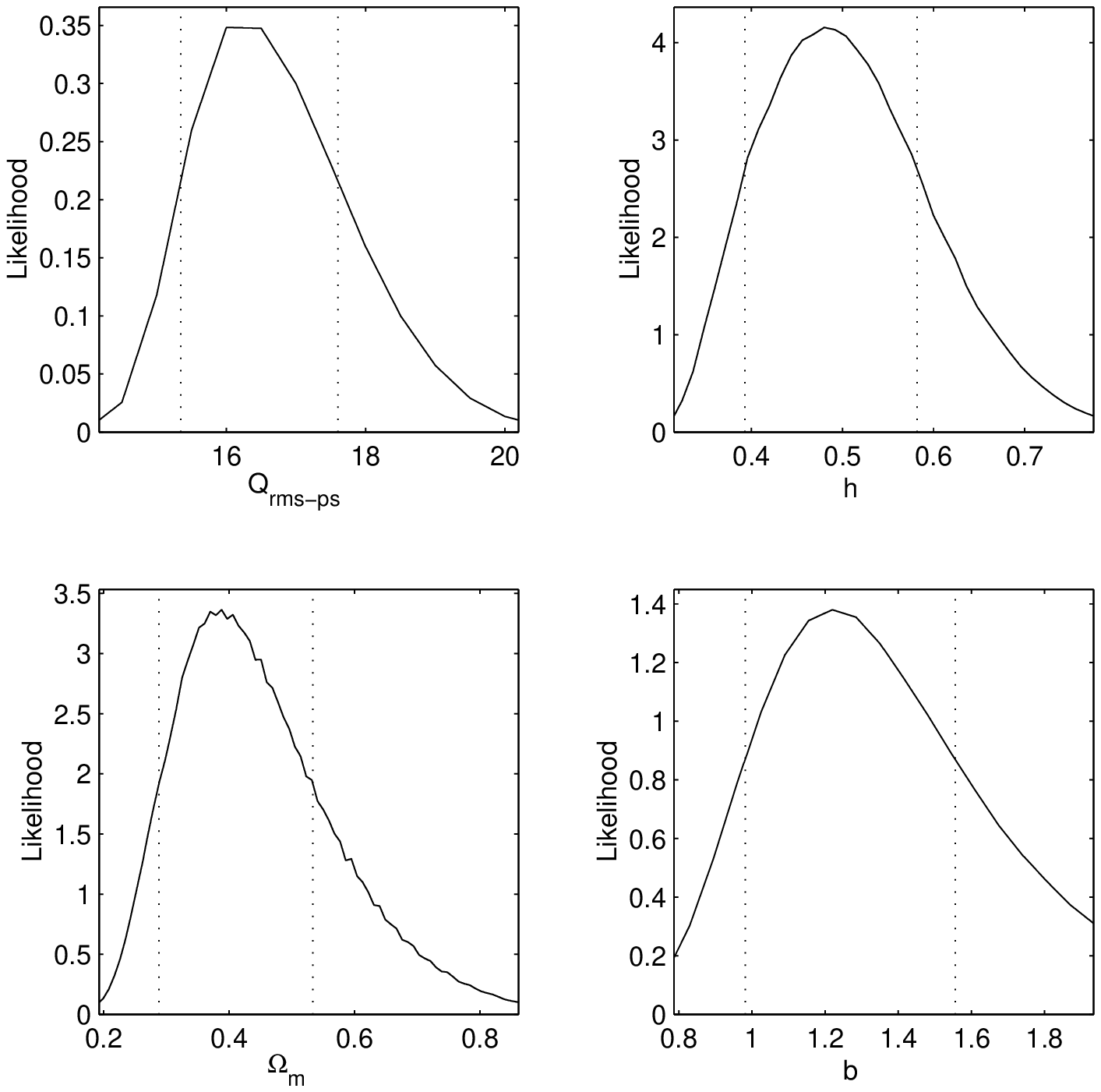,width=8cm}}
\caption{FIGURE 3 The probability distributions, using CMB and
IRAS data, after marginalising over the other free parameters.}
\end{figure}
The best fit results from the joint analysis of the two data sets
on all the free parameters are shown in
Table~1, for which we can derive $\Omega_b=0.085$, $\sigma_8=0.67$ and
shape parameter (Sugiyama 1995, Efstathiou, Bond \& White 1992) $\Gamma=0.15$. A 
detailed discussion of these estimates and comparison with other
results is contained in Webster \etal~1998, but
in broad terms it is clear that fairly sensible values have resulted,
which is encouraging for future prospects within this area. 
\vskip0.2cm
\centerline{\vbox{
{\footnotesize
Table~1. --- Parameter values at the joint optimum.
The 68\% confidence limits are shown,
calculated for each parameter by marginalising the likelihood over the other 
variables.\label{valuetable}}\\\\
\begin{tabular}{@{}lc r@{.}l @{$\,<\,$} c @{$\,<\,$} r@{.}l}
$\Omega_m$    &$0.39$    &  $0$ & $29$ &$\Omega_m$&$0$&$53$\\
$h$          &$0.53$    &  $0$ & $39$ &$h$      &$0$&$58$ \\
$Q$ ($\mu$K) &$16.95$   &  $15$& $34$ &$Q$      &$17$&$60$\\
$b_{\rm{IRAS}}$        &$1.21$    &  $0$ & $98$ &$b_{\rm{IRAS}}$    &$1$&$56$\\
\\
\end{tabular}
}}
For a spatially flat
model, the age of the universe is given by:
\begin{displaymath}
t = \frac{2}{3 H_0} \frac{\tanh^{-1}\sqrt{\Oml}}{\sqrt{\Oml}}
\end{displaymath}
which evaluates to 16.5 Gyr in the current case, again compatible
with previous estimates.

\bsk
\ni 4. COMBINING CMB, SUPERNOVAE AND LSS DATA
\ssk
\ni 

Finally, we combine all three data sets under the assumption applied in
the above CMB+IRAS analysis and find that the data sets agree well and
tighten the constraints on the four parameters investigated.
\begin{figure}[!t]
\centerline{\psfig{file=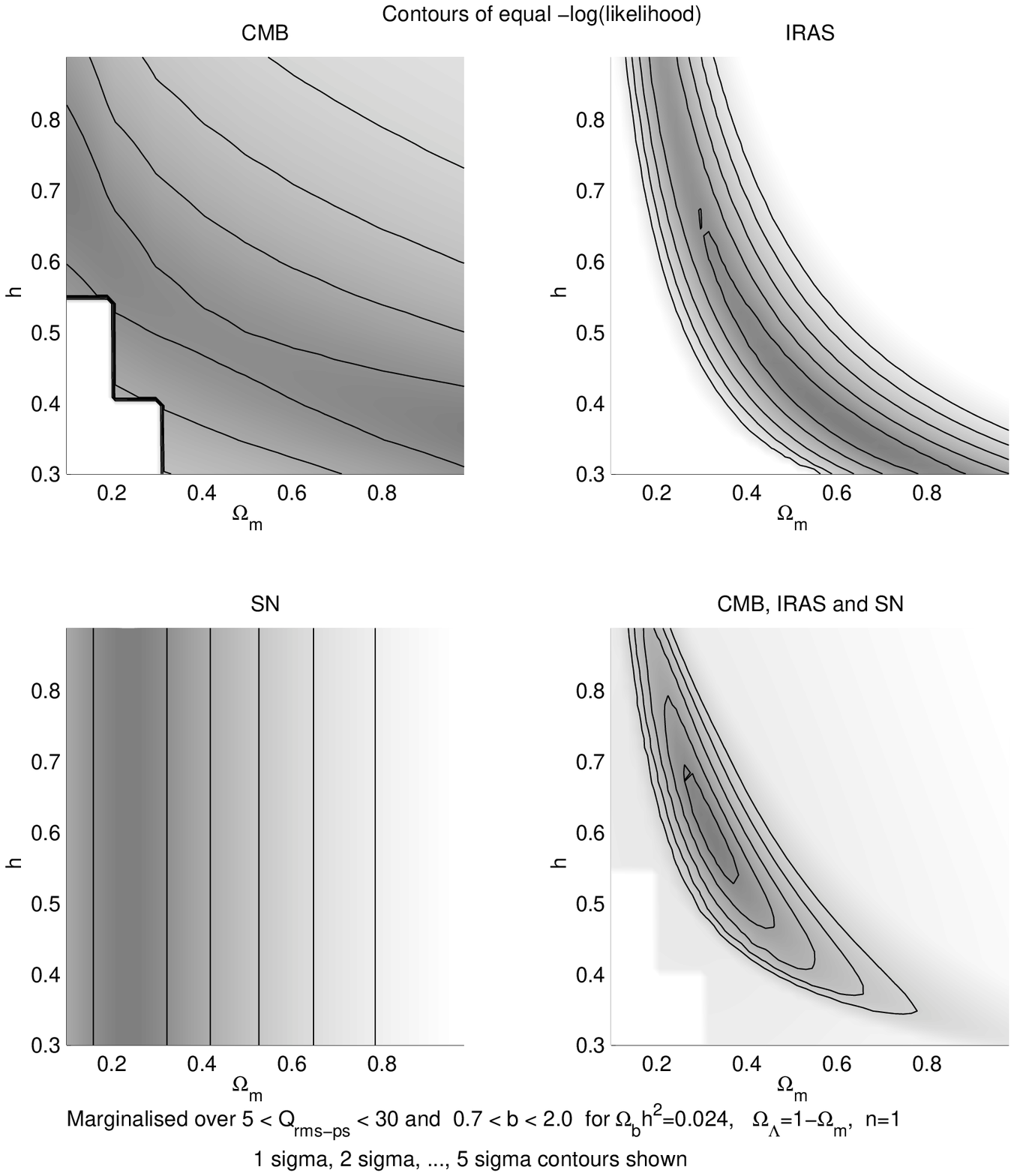,width=12cm}}
\caption{FIGURE 4 Likelihoods in the $h$, $\Omega_m$ plane after
marginalising over $Q_{\rm{rms-ps}}$ and $b_{\rm{IRAS}}$ for CMB,
IRAS, SN and all three data sets combined.
} 
\end{figure}
We find that $Q_{\rm{rms-ps}}=17.5 \pm 1 \mu$K, $H_0=59 \pm 8$ $\rm{km s^{-1}
Mpc^{-1}}$, $\Omega_m=0.33 \pm 0.07$ and $b_{\rm{IRAS}}=1.05 \pm 0.2$,
which may be compared to the results in Table 1. For a flat Universe,
as considered here, the SN results shift $h$ up and $\Omega_m$
down slightly compared to the results found using CMB and IRAS data
alone.

\bsk
\ni 5. CONCLUSIONS
\ssk
\ni 

We have combined CMB, LSS and SN data sets and found good agreement
between them, despite the restrictive nature of our assumptions. Due
to the complementary nature of the data sets the error bars on the
cosmological parameters are dramatically reduced by making such
combinations.

\bsk
\baselineskip = 12pt
{\abstract \ni ACKNOWLEDGMENTS
We would like to acknowledge collaboration with Ofer Lahav (Racah
Institute of Physics, The Hebrew University, Jerusalem, IoA
Cambridge), George Efstathiou and Matthew Webster (IoA Cambridge) and
Graca Rocha (Kansas). SLB acknowledges a PPARC studentship.

}

\bsk
\baselineskip = 12pt


{\references \ni REFERENCES
\ssk

\ref
{Bond}, J.~R. and {Jaffe}, A.~H. 1998, preprint (astro-ph/9809043).
\ref
{Efstathiou}, G., {Bond}, J.~R., and {White}, S. D.~M. 1992, \mnras,
258, 1.
\ref
Fisher, K. and Nusser, A. 1996, \mnras, 279, L1.
\ref
{Fisher}, K.~B., {Scharf}, C.~A., and {Lahav}, O. 1994, \mnras, 266, 219
\ref
Gawiser, E., \&  Silk, J. 1998, Science, 280, 1405
\ref
{Hancock}, S., {Rocha}, G., {Lasenby}, A.~N., and {Gutierrez}, C.~M. 1998,
\mnras, 294, L1
\ref
Heavens, A. and Taylor, A. 1995, \mnras, 275, 483
\ref
Lineweaver, C. 1998, \apj, in press (astro-ph/9805326)
\ref
Lineweaver, C., Barbosa, D., Blanchard, A., and Bartlett, J. 1997,
Astron.Astrophys., 322, 365
\ref
Perlmutter, S. et al. 1998
Poster displayed at the American Astronomical Society meeting in
Washington, D.C., January 9, 1998, 
\\ http://panisse.lbl.gov/public/papers/aasposter198dir/aaasposter.html.
\ref
{Seljak}, U. and {Zaldarriaga}, M. 1996, \apj, 469, 437
\ref
{Sugiyama}, N. 1995 \apjs, 100, 281
\ref
Tegmark, M. 1998, \apjl, submitted (astro-ph/9809201)
\ref
Tegmark, M., Eisenstein, D., Hu, W., and Kron, R. 1998,
\apj, submitted (astro-ph/9805117)
\ref
{Tytler}, D., {Fan}, X.~M., and {Burles}, S. 1996, \nat, 381, 207
\ref
{Webster}, A.~M., {Bridle}, S.~L., {Hobson}, M.~P., {Lasenby}, A.~N., {Lahav},
  O., and {Rocha}, G. 1998, \apjl, submitted (astro-ph/9802109)
\ref
Willick, J., Strauss, M., Dekel, A., and Kolatt, T. 1997, \apj, 486, 629
}                      

\end{document}